\documentclass[preprintnumbers,amsmath,amssymb,pre,floatfix,twocolumn,superscriptaddress]{revtex4} 
\usepackage{lipsum}
\usepackage{graphicx}
\usepackage{bm}
\usepackage{enumerate}
\usepackage[utf8]{inputenc}
\usepackage[usenames]{color}

\begin{document}

\title{Stability windows for proto-quark stars}

\author{V. Dexheimer}
\email{vdexheim@kent.edu}
\affiliation{Depto de F\'{\i}sica - CFM - Universidade Federal de Santa Catarina  Florian\'opolis - SC - CP. 476 - CEP 88.040 - 900 - Brazil}
\affiliation{Department of Physics, Kent State University, Kent OH 44242, USA}

\author{J. R. Torres}
\email{james.r.torres@posgrad.ufsc.br}
\affiliation{Depto de F\'{\i}sica - CFM - Universidade Federal de Santa Catarina  Florian\'opolis - SC - CP. 476 - CEP 88.040 - 900 - Brazil}

\author{D. P. Menezes}
\email{debora.p.m@ufsc.br}
\affiliation{Depto de F\'{\i}sica - CFM - Universidade Federal de Santa Catarina  Florian\'opolis - SC - CP. 476 - CEP 88.040 - 900 - Brazil}

\date{\today}

\begin{abstract}
We investigate the existence of possible stable strange matter and related stability windows at finite temperature for different models that are generally applied to describe quark stars, namely, the quark-mass density dependent model, the MIT bag model and the Nambu-Jona-Lasinio model. We emphasize that, although the limits for stable strange matter depend on a comparison with the ground state of $^{56}Fe$, which is a zero temperature state, the quantity that has to be used in the search for strange matter in proto-quark stars is the free energy and we analyze stability windows up to temperatures of the order of 40 MeV. The effects of strong magnetic fields on stability windows are computed and the resulting mass-radius relations for different stages of the proto-quark star are analyzed.

\begin{description}
\item[PACS numbers: 12.39.-x,26.60.-c,95.30.Tg]
\end{description}

\end{abstract}

\maketitle

\section{Introduction}

Neutron stars are believed to be the remnants of supernova explosions \cite{Lattimer2004,prak97,Glen00}. They are born hot and rich in leptons. During the very beginning of the evolution, an energy of the order of $10^{53}$ ergs is radiated away from the neutron star by neutrinos, in a process known as deleptonization. Theoretical studies involving different possible equations of state that result in different matter composition have to be performed. This is because the temporal evolution of the star in the so-called Kelvin-Helmholtz epoch, during which the remnant compact object changes from a hot and lepton-rich to a cold and deleptonized star, depends on two key ingredients: the equation of state (EoS) and its associated neutrino opacity at supranuclear densities \cite{pons1999,pons2001,marcelo}. 

One class of EOS assumes that the neutron star is composed only of hadrons, plus an essential small 
admixture of electrons. In order to construct appropriate EOS's for these hadronic stars, one must rely on models which describe nuclear matter bulk properties. The interior of the neutron star can also include quarks and, in this case, two possibilities have been investigated: hybrid stars, constituted by a hadronic phase and a quark 
core, or composed only by charge neutral quark matter in $\beta$-equilibrium. In hybrid stars, low density regions are composed of hadronic matter, but in high density regions deconfinement of the quarks occurs, leading to the formation of a quark phase. Many different EOS's have been built, including EOS's for the phase immediately after the formation of the neutron star, when the neutrinos are still trapped, and for the subsequent (deleptonized) phases after the neutrinos escape.

The assumption underlying the existence of quark stars is based on the Bodmer-Witten conjecture \cite{Itoh,bodmer, witten}. These authors have claimed that it is possible that the interior of a neutron-like star does not consist primarily of hadrons, but rather of strange matter (SM), which is composed of deconfined quarks, including up, down 
and strange quarks, plus the leptons necessary to ensure charge neutrality  and $\beta$-equilibrium. This possibility arises because a phase transition from hadronic to quark phase is possible at densities present in the interior of neutron stars. It has been argued \cite{olinto} that strange matter is the true ground state of all matter. If this is the case,  as soon as the core of the star converts to the quark phase, the entire star converts. SM was first considered in calculations obtained within the MIT bag model framework \cite{mit1,mit2}. More sophisticated treatments for SM, based on the Nanbu-Jona-Lasinio \cite{njl,njl2,hybrid2,hybrid4,hybrid6}, the color flavor locked phase 
\cite{Alford:2007xm,sergio1,sergio2,lugones2010,laura} and the quark-mass density dependent \cite{fowler} models also exist in the literature. However, the central densities inside a compact object are not known
and its inner composition could well be one of the superconducting phases (the
two-superconduction color phase (2SC)  or the CFL phase) \cite{sedrakian2011}, but this is still a source of speculation. 
Hence, investigating the quark phase without taking into account the CFL phase is indeed important and
our choice of models for the present work is based on the assumption that the
CFL phase transition takes place at higher densities than the deconfinement phase transition \cite{laura}. 
In the present paper, we restrict ourselves to constant chiral condensates, while an improved discussion should take into account particle-hole inhomogeneous condensates, which can become energetically favorable under certain densities and physical conditions \cite{novos,novos2,novos3}.

An important ingredient in the SM hypothesis is the stability window, identified with the model parameters that are consistent with the fact that two-flavor quark matter (2QM) must be unstable (i.e., at zero temperature its energy per baryon has to be larger than 930 MeV, the iron binding energy) and SM (three-flavor quark matter) must be stable (i.e., its energy per baryon must be lower than 930 MeV, also at $T=0$). This problem has already been discussed in the literature, but we believe that some aspects deserve more investigation and this is the main goal of the present work.
First of all, at zero temperature, the stability window is normally obtained for neutral matter in $\beta$-equilibrium, as for instance in Ref.~\cite{lugones95,1995NuPhA.588..365T,Prakash:1996xs}. Nevertheless, one has to bear in mind that stable nuclear matter (as in iron) is not charge neutral and does not contain electrons. Actually, its proton fraction is $Y_p=0.46$, very close to symmetric matter and this is a good reason to analyze also matter with equal quark chemical potentials, as done in Ref.~\cite{james} for zero temperature systems. Moreover,  an important characteristic of the deconfinement transition is that deconfined quark matter is transitorily out of equilibrium with respect to weak interactions \cite{nucleation1,nucleation2,lugones2010}. It was also shown in Ref.~\cite{james} that slightly different results are obtained if one considers matter with identical quark chemical potentials, corresponding to equal quantities of $u$ and $d$ quarks in two-flavor-matter, or charge neutral matter in $\beta$-equilibrium, as expected in stars. For the models under investigation, the stable parameter region is larger if the $\beta$-equilibrium condition and charge neutrality are imposed. 

Another point worth mentioning refers to finite size effects, very often disregarded in the literature. As shown in Ref.~\cite{mit2}, 
 if finite size effects are taken into account, iron binding energy is around 4 MeV lower, i.e. 934 MeV.  However, one can see \cite{james} that
even if finite size effects are taken into account, the stability window remains practically unchanged.

Understanding the stability of strange matter in proto-quark stars is also important because it is known that stars cool down slowly after deleptonization. According to numerical simulations, during the first tens of seconds of evolution the proto-neutron star cools from $T=40$ MeV (in the center) to temperatures below 2–4 MeV . One does not expect that proto-quark stars are constant in temperature. Actually, it is believed that the entropy per baryon is fixed, so that the temperature varies in the interior of the star, according to its density. A reasonable assumption is that during the evolution process, different snapshots can be simulated though different entropies per baryon and trapped neutrinos. See Refs.~\cite{Dexheimer:2008ax,veronica} for some examples of studies of isolated star evolution. This analysis is important as it has been shown that evolution of isolated stars differs from the one of stars in binary systems throughout the whole star life \cite{Li:2013xv,Li:2007kg}.

The stability window at finite temperature has already been discussed in the past \cite{chmaj89,chakrabarty93, lugones95t,su2002}. In all these papers, except in Ref.~\cite{chmaj89}, the importance of the free energy was disregarded, but even in Ref.~\cite{chmaj89}, where it was introduced in the calculations, the quantity used in the search for stable strange matter was always the binding energy per baryon. Of course, as expected from calculations in the macrocanonical or grand-canonical ensemble, the quantity related to the thermodynamical potential is the free energy per baryon (${\cal F}/A = f/\rho_B = (\epsilon - Ts)/\rho_B$), where $f$ is the free energy density, $\rho_B$ the baryon density, $\epsilon$ the energy density, $T$ the temperature and $s$ the entropy density of the system \cite{greiner}. For zero temperature systems, the free energy density becomes the energy density and hence, the binding energy per baryon ($B/A=\epsilon/\rho_B$) is analyzed at zero temperature in the search for stable 
matter.
The choice of appropriate parameters compatible with stable SM at finite temperature systems requires, in addition to the 
investigation performed at zero temperature, a careful study of the free energy per baryon. A very similar situation is seen when one looks for the possible existence of the pasta phase in finite systems \cite{pastat1,pastat2}.
  Of course, the lower limits of the stability windows, given by the points where the
two-flavor quark matter has an energy per baryon larger than 930 MeV, remain the same because the 
$^{56}Fe$ ground state is a zero temperature system.
 
Another important aspect related to neutron stars is their magnetic fields. While magnetars are expected to bear extremely high magnetic fields of up to $B=10^{15}$ G on the surface \cite{c11,c12,c13,c14}, all pulsars have magnetic fields of some strength \cite{b1}. For this reason, any complete analysis of pulsar features, including stability windows, should take into account magnetic field effects. Note that whenever tackling this problem, anisotropy is a matter that has to be carefully considered \cite{laura,Chaichian:1999gd,Martinez:2003dz,PerezMartinez:2005av,Felipe:2007vb%
,PerezMartinez:2007kw,Ferrer:2010wz,Orsaria:2010xx,veronica,Isayev:2011ug,Isayev:2013sq,Strickland:2012vu,Sinha:2012cx}, once it may be important for ultra high magnetic fields. In this work, however, we are not going to take into account anisotropy of matter, since we are not primarily interested in analyzing macroscopic properties of stars, but general trends of different microscopic models in specific situations.

For the above mentioned reasons, it is appropriate to have a better understanding of stability windows also at finite temperature (including the case of fixed entropy per baryon) subject to magnetic fields. This is the scope of the present
paper, organized as follows: in Section II, we briefly discuss the formalism and the EOS's of different models; in Section III we present the results and conclusions and in the last section, the final remarks are drawn.

\section{Formalism}

We discuss next the following models for proto-quark stars: the quark-mass density dependent (QMDD) model, the MIT bag model and the Nanbu-Jona-Lasinio (NJL) model. We restrict ourselves to the main formulae and refer the reader to previous references for detailed calculations.

\subsection{QMDD model}

We start from the QMDD model \cite{fowler,chakrabarty91, james,lugones95,wang00,peng00,su2008}, which is based on a phenomenological approach where the dynamical masses of the three lightest quarks scale inversely with the baryon number density
\begin{equation}
m_{u,\overline{u}}^{\ast }=m_{d,\overline{d}}^{\ast }=\frac{C}{3\rho_B},~~~~ m_{s,\overline{s}}^{\ast }=m_{s,\overline{s}}+\frac{C}{3\rho_B} \mbox{,}
\label{ansatz_massa}
\end{equation}
where $C$ is the constant energy density in the zero quark density limit.

The pressure, energy density and baryonic density of the system are, respectively, given by
\begin{equation}
p=\sum_{i}\frac{\gamma_{i}}{3}\int \frac{d^{3}k_i}{\left(2\pi \right)^3}\frac{k_i^{2}}{\sqrt{k_i^{2}
+m_{i}^{\ast2}}}\left(  {f_{+ i}}+{f_{- i}} \right)-\mathcal{B}(\rho_B,f_{\pm i}) \mbox{,}
\label{pressao_qmdd}
\end{equation}
\begin{equation}
\epsilon=\sum_{i}\gamma_{i}\int\frac{d^{3}k_i}{\left(2\pi \right)^3}\sqrt{k_i^{2}+m_{i}^{\ast2}}\left({f_{+ i}}+{f{_-,i}} \right)+\mathcal{B}(\rho_B,f_{\pm i}) \mbox{,}
\label{densidade_energia_qmdd}
\end{equation}
\begin{equation}
\rho_B=\sum_i \frac{\rho_{i}}{3}=\sum_i \frac{\gamma_{i}}{3}\int \frac{d^{3}k}{\left(2\pi \right)^3}\left({f_{+,i}}-{f_{-,i}} \right) \mbox{,}
\label{densidade_quarks_qmdd}
\end{equation}
where
\begin{eqnarray}
\mathcal{B}(\rho_B,f_{\pm})&=&\sum_{i}\gamma_{i}\int \frac{d^{3}k_i}{\left(2\pi \right)^3}\frac{m_{i}^{\ast}}{\sqrt{k_i^{2}+m_{i}^{\ast2}}}\nonumber\\
&\times&\left( \frac{C}{3\rho_B}\right)\left({f_{+ i}}+{f_{- i}} \right) \mbox{,}
\label{termo_de_confinamento}
\end{eqnarray}
and $\gamma_i$ is the degeneracy of each quark $i=u,d,s$ taking into account spin and number of colors. The distribution function for quarks ($f_+$) and antiquarks ($f_-$) are the Fermi-Dirac distributions
$f_{\pm i}=\left[ 1+\exp \left[ \left( E_{i}^{\ast } \mp \mu _{i}\right)/T\right] \right]^{-1}$ where $\mu_i$ is the chemical potential of each particle species and $E_{i}^{\ast }\left( p\right) =\sqrt{k_i^{2}+m_{i}^{\ast 2}}$.

This model presents a thermodynamical inconsistency which was tackled in different ways in the literature and resulted in different versions of the QMDD model. This problem was discussed in many papers \cite{wang00,peng00,su2008,james} and we restrict ourselves to two versions: 1) the  most commonly used one \cite{lugones95}, where the pressure at the density corresponding to the minimum of the free energy per baryon could be non-zero, depending on the matter studied (SM or 2QM) and to which we refer next as version 1 (QMDDv1)
 and 2) the one that presents a remedy to the thermodynamical inconsistency, in such a way that the minimum of
the energy per baryon corresponds to the point of zero pressure \cite{peng00}, and to which we refer as version 2 (QMDDv2).
The confinement parameter values for which the stability windows are obtained in different versions of the model at zero temperature are different, as shown in Ref.~\cite{james}.

\subsection{MIT bag model}

For the MIT bag model \cite{mit1,mit2} the quark masses are fixed. Therefore, expressions (\ref{pressao_qmdd}), 
(\ref{densidade_energia_qmdd}) and (\ref{densidade_quarks_qmdd}) have
$m^{\ast }$ replaced by $m$ and $\mathcal{B}(\rho_B,f_{\pm i})$ by a bag constant $\mathcal{B}$. We take advantage of the simplicity of this model and include magnetic field effects in the calculation. In this case, the main expressions
are written next (see Ref.~\cite{veronica} for details). For the pressure, energy density and baryonic density they read, respectively
\begin{equation}
p =\sum_i \sum_\nu \frac{\gamma_i}{2 \pi^2}|q_i|B \int dk_i \frac{k_i^2}{\sqrt{k_i^2+\bar{m_i}^2}} ({f_{+ i}}+{f_{- i}}) -\mathcal{B} \mbox{,}
\label{pbag}
\end{equation}
\begin{equation}
\epsilon=\sum_i \sum_\nu \frac{\gamma_i}{2 \pi^2}|q_i|B \int dk_i \sqrt{k_i^2+\bar{m_i}^2} ({f_{+ i}}+{f_{- i}}) +\mathcal{B} \mbox{,}
\label{ebag}
\end{equation}
\begin{equation}
\rho_B=\sum_i \frac{\rho_{i}}{3}= \frac{1}{3} \sum_i \sum_\nu \frac{\gamma_i}{2 \pi^2}|q_i|B \int dk_i ({f_{+ i}}-{f_{- i}}) \mbox{,}
\label{rhobag}
\end{equation}
where $\gamma_i$ is once more the degeneracy of each quark $i=u,d,s$ taking into account spin and number of colors. The difference now is that both spin projections contribute for Landau levels $\nu>0$, but only one of them contributes for $\nu=0$. 
The sum in the Landau levels runs until the last one is filled (strictly, it goes up to
infinity). Numerically, we have developed a robust way to check for the
convergence of the results. When they no longer change with a very
good precision, we are certain that the last possible level for a fixed magnetic field was filled. The same procedure is valid
for all models with magnetic fields at finite temperature because only at zero temperature the maximum Landau level has a definite
value.

Note that we only consider the case where the magnetic field is parallel to the $z$ direction so the energy levels in the $x$ and $y$ directions are quantized. In this case $q_i$ is the electric charge of each quark, $B$ is the magnetic field strength, $\bar{m_i}=\sqrt{m_i^2+2|q_i|B\nu}$ the mass of each quark ($m_{u,d}=5$ MeV, $m_s$ can take different values) modified by the magnetic field and $E_i=\sqrt{k_i^2+\bar{m_i}^2}$.

\subsection{NJL model}

For the NJL model (see Ref.~\cite{njl,njl2} for details), we need to evaluate the grand-canonical thermodynamical potential for the three-flavor quark sector
\begin{equation}
\Omega=-P=\epsilon-T s-\sum_i\mu_i\rho_i-\Omega_0,
\label{thermopot}
\end{equation}
where $\Omega_0$ ensures that $\Omega=0$ in the vacuum. As a result, we obtain for the pressure and entropy density of the system in the mean field approximation in the presence of a magnetic field the expressions given next. See the appendix for the formulas without magnetic field effects and refer
to Refs.~\cite{njlb1,njlb2,njlb3} for details on the calculations of the expressions given below:
\begin{equation}
P=\theta_u+\theta_d+\theta_s 
-2G(\phi_u^2+\phi_d^2+\phi_s^2) + 4K \phi_u \phi_d \phi_s \mbox{,}
\end{equation}
\begin{eqnarray}
s&=&-\sum_i\sum_\nu\frac{\alpha_\nu N_c |q_i|B }{4 \pi^2}\int dk_i \Big[f_{+i} \ln\left(f_{+i}\right)
+\left(1-f_{+ i}\right)  \nonumber\\
&\times&\ln\left(1-f_{+ i}\right)+f_{- i} \ln\left(f_{- i}\right)
+\left(1-f_{- i}\right) \ln\left(1-f_{- i}\right)\Big] \mbox{.}\nonumber\\ 
\label{entropyb}
\end{eqnarray}
The energy density of the system can be calculated from Eq. (\ref{thermopot}) taking the baryon density to be the same as in Eq.(\ref{rhobag}). For this model we split the degeneracy of each quark into the spin degeneracy $\alpha_\nu$ (again both projections contribute for $\nu>0$ but only one for $\nu=0$) and color degeneracy $N_c$. In the above equations, $G$ and $K$ are coupling constants, $q_i$ is the electric charge of each quark, $B$ is the magnetic field strength
and the contribution from the gas of quasiparticles for each flavor $\theta_i=\left(\theta^{vac}_i+\theta^{mag}_i+\theta^{med}_i\right)_{M_i}$ contains three different contributions: the vacuum, the magnetic and the medium one given by
\begin{equation}
\theta^{vac}_i=- \frac{N_c}{8\pi^2} \left \{ M_i^4 \ln \left[\frac{(\Lambda+ \epsilon_\Lambda)}{M_i} \right]
-\epsilon_\Lambda \, \Lambda\left(\Lambda^2 +  \epsilon_\Lambda^2 \right ) \right \} \mbox{,}
\end{equation}
\begin{equation}
\theta^{mag}_i= \frac {N_c (|q_i| \mathcal{B})^2}{2 \pi^2} \left [ \zeta^{(1,0)}(-1,x_i) -  \frac {1}{2}( x_i^2 - x_i) \ln x_i +\frac {x_i^2}{4} \right ] \mbox{,}
\end{equation}
\begin{eqnarray}
\theta^{med}_i&=&\sum_\nu\frac{\alpha_\nu N_c|q_i|B }{4\pi^2}
\int dk_i \Big\{\ln\left[1+\exp[-(E^*_i-\mu_i)/T]\right]\nonumber \\
&+&\ln\left[1+\exp[-(E^*_i + \mu_i)/T]\right] \Big\} \mbox{.}
\label{PmuB}
\end{eqnarray}
Above, we have defined $\epsilon_\Lambda=\sqrt{\Lambda^2 + M_i^2}$ with $\Lambda$ representing a non-covariant ultra violet cut off \cite{Ebert:1999ht},  $x_i = M_i^2/(2 |q_i| \mathcal{B})$ and $\zeta^{(1,0)}(-1,x_i)= d\zeta(z,x_i)/dz|_{z=-1}$ with $\zeta(z,x_i)$ being the Riemann-Hurwitz zeta function.

\begin{figure}[t!]
\centering
\includegraphics[width=0.45\textwidth]{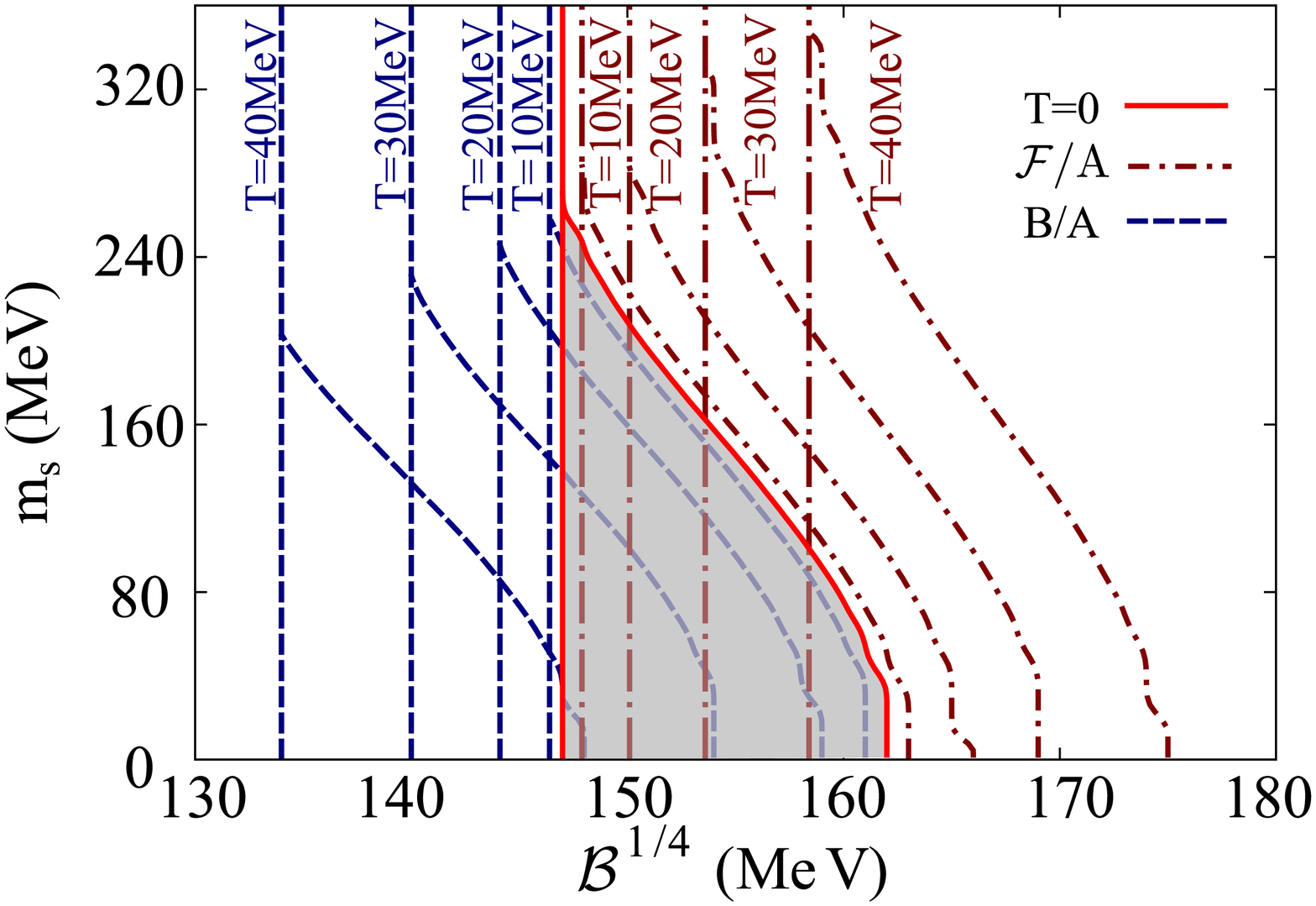}
\caption{(Color online) Stability windows of SM obtained with the MIT model through the analysis of the free energy per baryon and the binding energy per baryon (both $\eqsim 930$ MeV) shown for different temperatures.}
\label{fig_mit_energy_free_energy}
\end{figure}

Each of the quark condensates, $\phi_i=\langle{\bar q}_i q_i\rangle=(\phi_i^{vac}+\phi_i^{mag}+\phi_i^{med})_{M_i}$ also contains three different contributions: the vacuum, the magnetic and the medium one given by
\begin{eqnarray}
\phi_i^{vac} &=& -\frac{N_c  M_i}{2\pi^2} \left[\Lambda \epsilon_\Lambda-{M_i^2}\ln \left ( \frac{\Lambda+ \epsilon_\Lambda}{{M_i }} \right ) \right ] \mbox{,}
\end{eqnarray}
\begin{eqnarray}
\phi_i^{mag}&=& -\frac{N_c M_i |q_i| \mathcal{B} }{2\pi^2}\left [ \ln \Gamma(x_i) -\frac {1}{2} \ln (2\pi) \right . \nonumber \\
&+& \left . x_i -\frac{1}{2} \left ( 2 x_i-1 \right )\ln (x_i) \right ] \mbox{,}
\end{eqnarray}
\begin{eqnarray}
\phi_i^{med}&=&\sum_{\nu}\frac{\alpha_\nu N_c M_i |q_i| B}{2\pi^2}\int dk_i \frac{(f_{+ i} + f_{- i})}{E^*_i} \mbox{,}
\label{MmuB}
\end{eqnarray}
where $E_i^*=\sqrt{k_i^2+s_i(\nu,B)^2}$ and $s_i(\nu,B)=\sqrt{M_i^2+2|q_i|B\nu}$ is the constituent mass of each quark modified by the magnetic field. Once more we consider the case where the magnetic field is parallel to the $z$ direction so the energy levels in the $x$ and $y$ directions are quantized
as done in Refs.~\cite{njlb1,njlb2,njlb3}.
Minimizing the grand-canonical thermodynamical potential $\Omega$ with respect to $M_i$ 
leads to three gap equations
\begin{equation}
M_i\,=\,m_i\,-4G \phi_i+\,2 K \phi_j\phi_k \mbox{,}
\label{efmass}
\end{equation}
with cyclic permutations of  $i,\, j,\, k$.

\begin{figure}[t!]
\centering
 \includegraphics[width=0.45\textwidth]{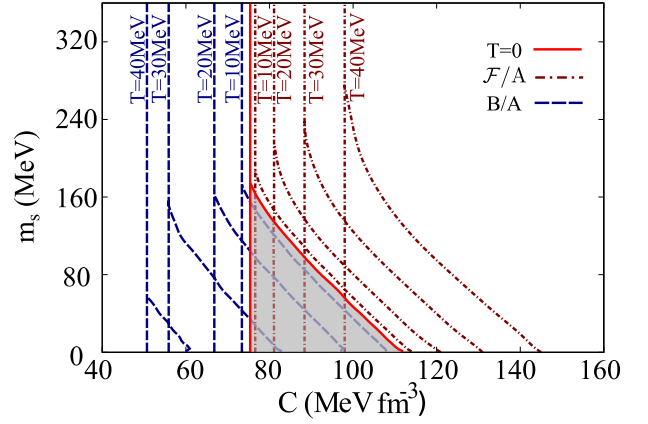}
\caption{(Color online) Stability windows of SM obtained with the QMDDv1 model through the analysis of the free energy per baryon and the binding energy per baryon (both $\eqsim 930$ MeV) shown for different temperatures.}
\label{fig_qmdd_energy_free_energy}
\end{figure}

\subsection{Chemical equilibrium}

In the description of compact stars, both charge neutrality and chemical equilibrium conditions have to be imposed  \cite{Glen00,quarkionicas1}. The first condition can be written for quarks and leptons as
\begin{equation}
2\rho _{u}=\rho _{d}+\rho _{s}+3\left( \rho _{e}+\rho _{\mu }\right) \mbox{,}
\end{equation}
and the second condition can be written as
\begin{equation}
\mu _{s}=\mu _{d}=\mu _{u}+\mu _{e}\mbox{, \ \ }\mu _{e}=\mu _{\mu } \mbox{,}
\label{qch}
\end{equation}
where the lepton densities can be calculated through Eq.~(\ref{densidade_quarks_qmdd}) or Eq.~(\ref{rhobag}), 
depending on the inclusion or not of the magnetic field, with appropriate substitutions for the masses and electric charge.
The lepton masses are $m_e=0.511$ MeV and $m_\mu=105.66$ MeV. For the electron and muon pressure and energy density we use Eqs. (\ref{pressao_qmdd}), and (\ref{densidade_energia_qmdd}) or (\ref{pbag}) and (\ref{ebag}), in the case without and with the inclusion of the magnetic field, respectively, all with $\mathcal{B}=0$.

In earlier stages, when the neutrinos are still trapped in the interior of the star, Eq.~(\ref{qch}) is replaced by
\begin{equation}
\mu_s=\mu_d=\mu_u+\mu_e -\mu_{\nu e} \mbox{,}
\label{qchtrap}
\end{equation}
\begin{equation}
\mu_\mu=\mu_e\ \ \ \ \text{and}\ \ \ \ \mu_{\nu_\mu}=\mu_{\nu_e} \mbox{.}
\end{equation}
The independent chemical potential $\mu_{\nu e}$ appears as the system gains one conserved quantity, the lepton fraction $ Y_l=(\rho_e+\rho_\mu+\rho_{\nu_e}+\rho_{\nu_\mu})/\rho_B$, which, according to simulations \cite{Burrows:1986me,marcelo}, can reach
$Y_l=0.4$. Note that the degeneracy factor for the neutrinos is $\gamma_i=1$ and, because they are uncharged, they follow Eq. \ref{pressao_qmdd}, \ref{densidade_energia_qmdd} and \ref{densidade_quarks_qmdd} (with $\mathcal{B}=0$), even in the presence of magnetic fields.

\begin{figure}[t!]
\centering
 \includegraphics[width=0.45\textwidth]{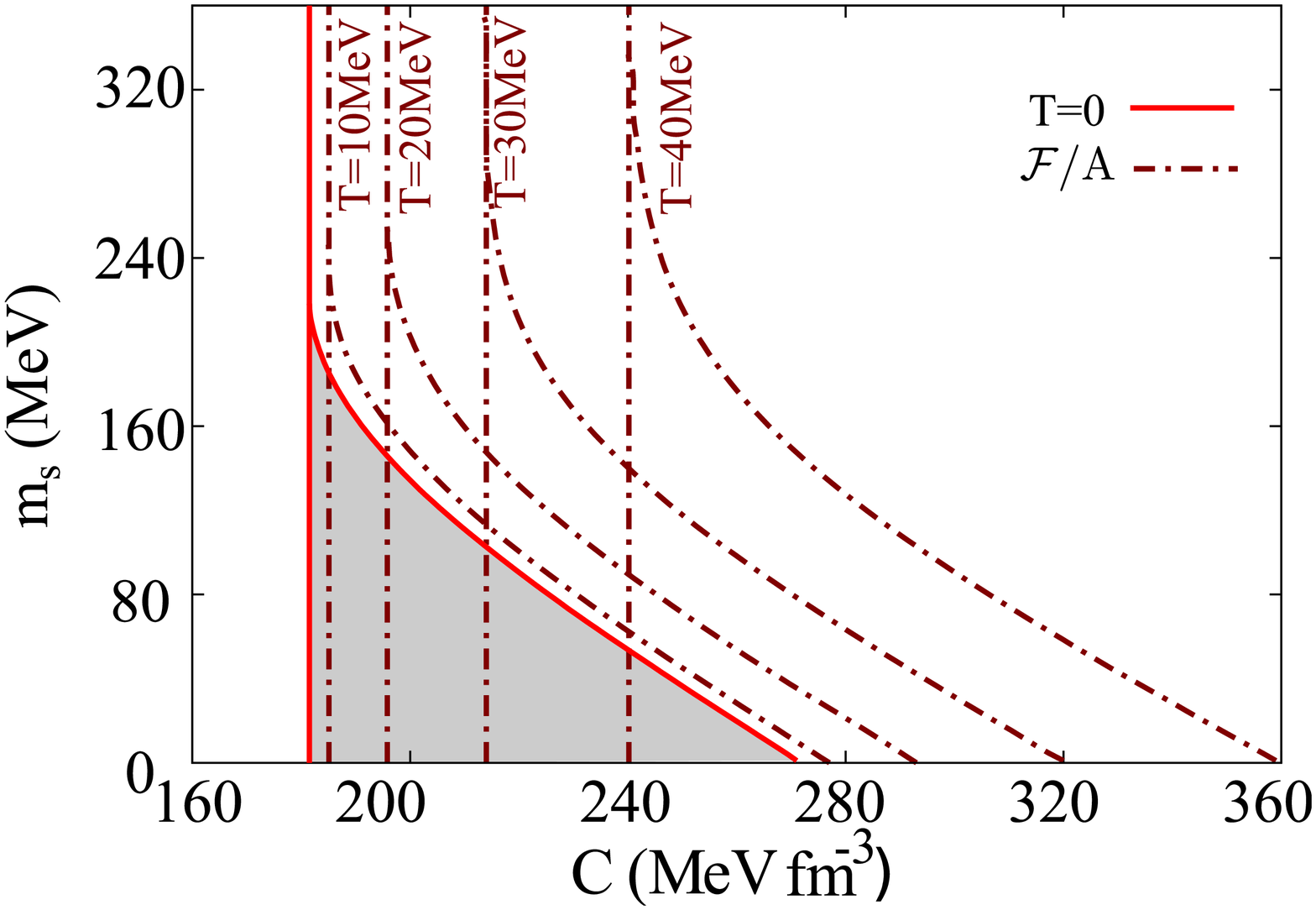}
\caption{(Color online) Stability windows of SM obtained with the QMDDv2 model through the analysis of the free energy per baryon 
shown for different temperatures.}
\label{fig_ddqm_energy_free_energy}
\end{figure}

\section{Results and Conclusions}

We start by analyzing the stability window related to proto-quark stars described by SM obtained with the MIT bag model. As stated in the Introduction, instead of considering the binding energy, the quantity that has to be studied in obtaining the upper limit of the stability window 
is the free energy. In order to show how different the results are, we plot both the binding energy per baryon and the free energy per baryon in Fig \ref{fig_mit_energy_free_energy} as a function of the bag constant and the strange quark mass for the MIT bag model, without the inclusion of magnetic field effects. Notice that we have used, for two-flavor quark matter (2QM), the fact that $\mu_u=\mu_d$, which gives symmetric matter ($\rho_u=\rho_d$) and, to be consistent, for SM we have used $\mu_u=\mu_d=\mu_s$. As the strange quark mass is much larger than the masses of quarks $u$ and $d$, its relative density is considerably lower. For $T=0$, the binding energy per baryon has been calculated numerically for 2QM and SM, respectively, and SM is stable in the shaded region shown in Fig.\ref{fig_mit_energy_free_energy}. The lower limit, vertical straight line, is due to the requirement that two-flavor quark matter is not absolutely stable. We have repeated the calculation for temperatures up to 40 MeV, as they 
are 
relevant for the proto-neutron star evolution simulations. Our results reproduce the same behavior as the ones given in Refs.~\cite{chmaj89,lugones95t,su2002} for the binding energy per baryon, i.e., as temperature increases, the stability windows move to the left, i.e., becomes more restrictive. However, when we consider the free energy per baryon, the stability window moves to the right and a wider range of constants becomes possible to ensure stable matter, contrary to what has been published so far.

\begin{figure}[t!]
\centering
\includegraphics[width=0.45\textwidth]{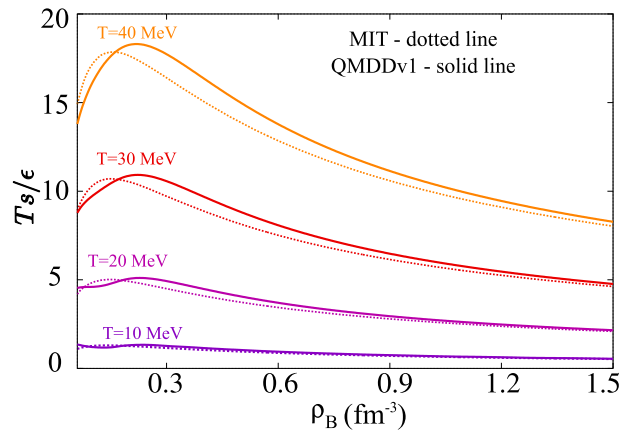}
\caption{(Color online) Ratio of the term $Ts$ and the energy density vs baryon density in SM. Shown for different models and different temperatures.}
\label{newfigure}
\end{figure}

In order to choose adequate values for the strange quark mass and the constant that enters in the quark masses in the QMDDv1 model, we display stability windows for finite temperatures in Fig. \ref{fig_qmdd_energy_free_energy}. The shaded area
corresponds once more to the zero temperature case. Note that considering the free energy per baryon as the criterion for stability, the windows move right with the increase of the temperature, as in the MIT model.
We analyze also the QMDDv2 model and the results are shown in Fig. \ref{fig_ddqm_energy_free_energy}.
One can see that the values of the confinement constants are somewhat different for this version of the model, with
obvious consequences on the resulting stellar matter configuration, as will be seen next.

\begin{table}[t!]
\begin{ruledtabular}
\begin{tabular}{ccccc}
 Set  & Model & $m_{s}$ (MeV) & $\mathcal{B}^{1/4}$ (MeV) & C (MeV$fm^{-3}$) \\ 
\hline
 A    & MIT   & $150$          & $155$                     &  N/A               \\ 
\hline
 B    & QMDDv1  & $150$          &   N/A                    & $78$             \\ 
 C    & QMDDv1  & $100$          &   N/A                      & $85$             \\ 
\hline
 D    & QMDDv2  & $150$          &    N/A                      & $194$             \\ 
 E    & QMDDv2  & $100$          &     N/A                     & $215$             \\ 
\end{tabular}
\end{ruledtabular}
\caption{\label{tab:table1}Parameters for the MIT and QMDD models.}
\end{table}

One should notice that at low temperatures, the binding energy and the free energy are indeed similar. 
However, one can see from Figs. \ref{fig_mit_energy_free_energy}  and  \ref{fig_qmdd_energy_free_energy}
 that at $T=20$ MeV, the discrepancy between both approaches are already non-negligible. 
For the MIT model, the stability window obtained with the energy
density ranges from $144.1 < B^{1/4} < 147.1$ MeV while the limit
becomes $147.1 < B^{1/4} < 150.1$ if the free energy density is
used. For the QMDDv1 model, the values move from $65.5 < C < 74.5$ to $74.5 < C < 79.9$ MeV fm$^{-3}$.
Moreover, if we consider the ratio of the term $Ts$ and the energy
density, we can see from Fig.\ref{newfigure} that it reaches almost 20 percent at $T=40$ MeV. 
A similar analysis, regarding the importance of investigating the free energy density,  was performed in \cite{wen2005}.

\begin{figure}[t!]
\centering
\includegraphics[width=0.45\textwidth]{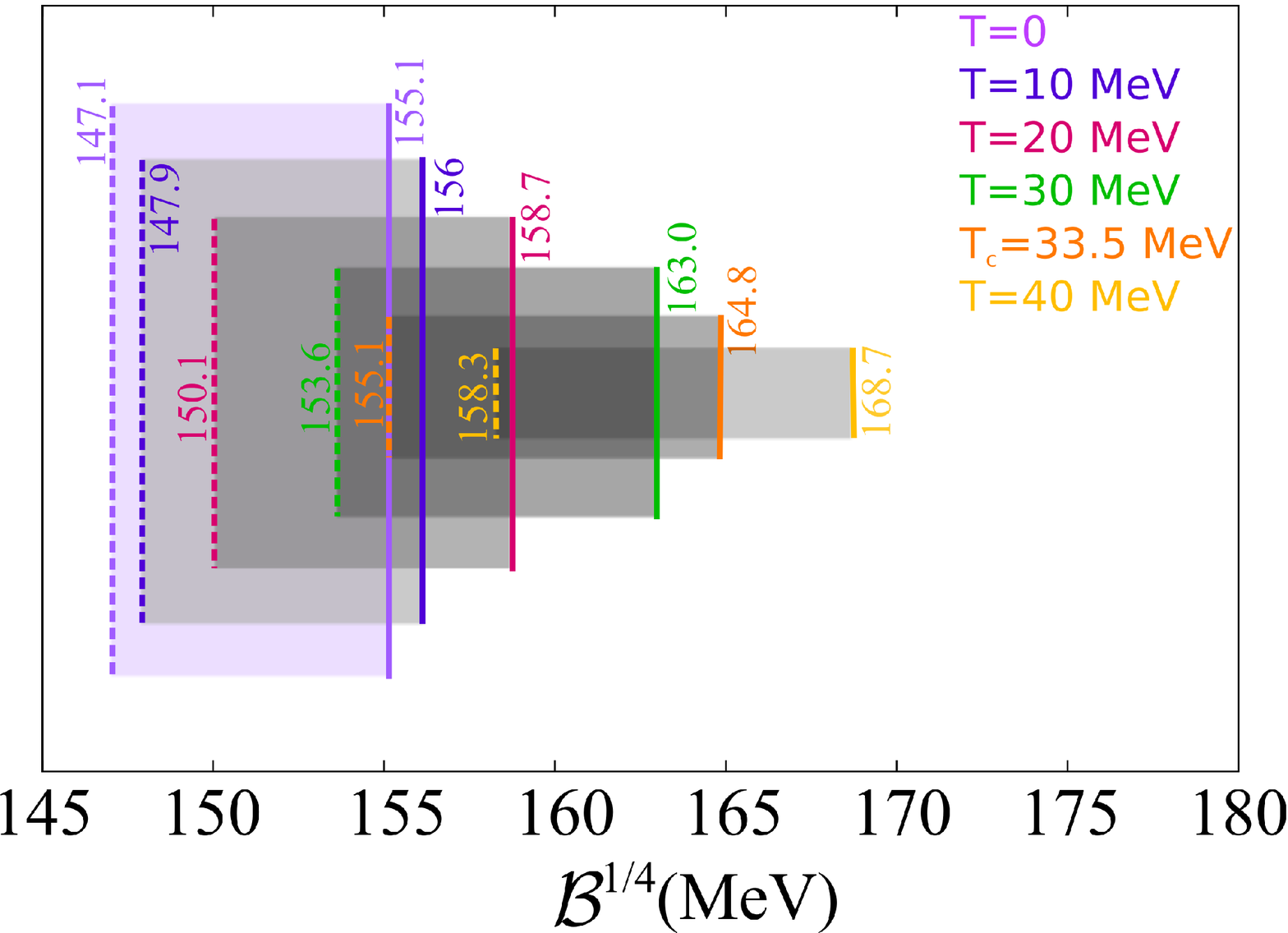}
\includegraphics[width=0.45\textwidth]{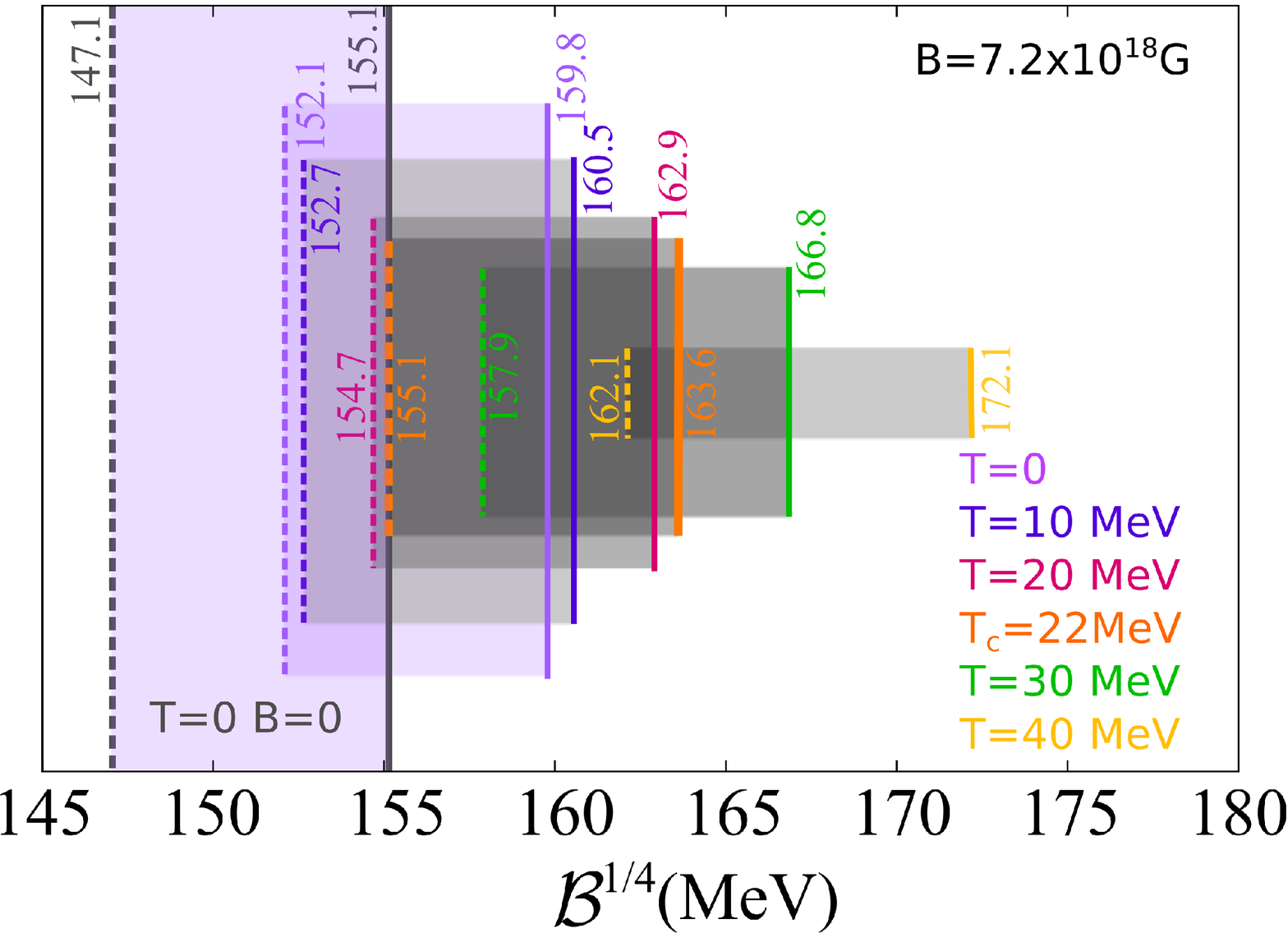}
\caption{(Color online) Stability windows of SM obtained with the MIT model with parametrization A without (top) and with (bottom) the inclusion of a magnetic field of $7.2\times10^{18}$ G shown for different temperatures. The orange rectangle shows the highest temperature that overlaps with $T=0$.}
\label{box_stability_mit}
\end{figure}

We then fix the strange quark mass to $150$ MeV (parametrization A in Table~\ref{tab:table1}) and calculate the overlap of stability windows at various temperatures through the analysis of
the free energy per baryon for the bag model. In Fig.~\ref{box_stability_mit} we display results without and with the inclusion of the magnetic field. For each temperature, 
we have used the prescription given above for unstable 2QM and stable SM to restrict the bag model parameter values. We can see that after a certain temperature, i.e., 
$T=33.5$ MeV for non-magnetized matter and $T=22$ MeV for magnetized matter, the boxes representing the parameter values stop overlapping. However, considering that 
the lowest limit should always be established for the $T=0/ B=0$ case, a wide range of parameters can be used for finite temperature matter inside stars. For example, 
if we consider matter with fixed entropy per baryon $S/A=2$, the temperature can reach 40 MeV in the center of the proto-quark star. In this case, $\mathcal{B}
^{1/4}$ can vary from $147$ 
to $168.7$ MeV, for unmagnetized stars and $147 < \mathcal{B}^{1/4} < 172.2$ MeV for magnetized stars.

\begin{figure}[t!]
\centering
 \includegraphics[width=0.45\textwidth]{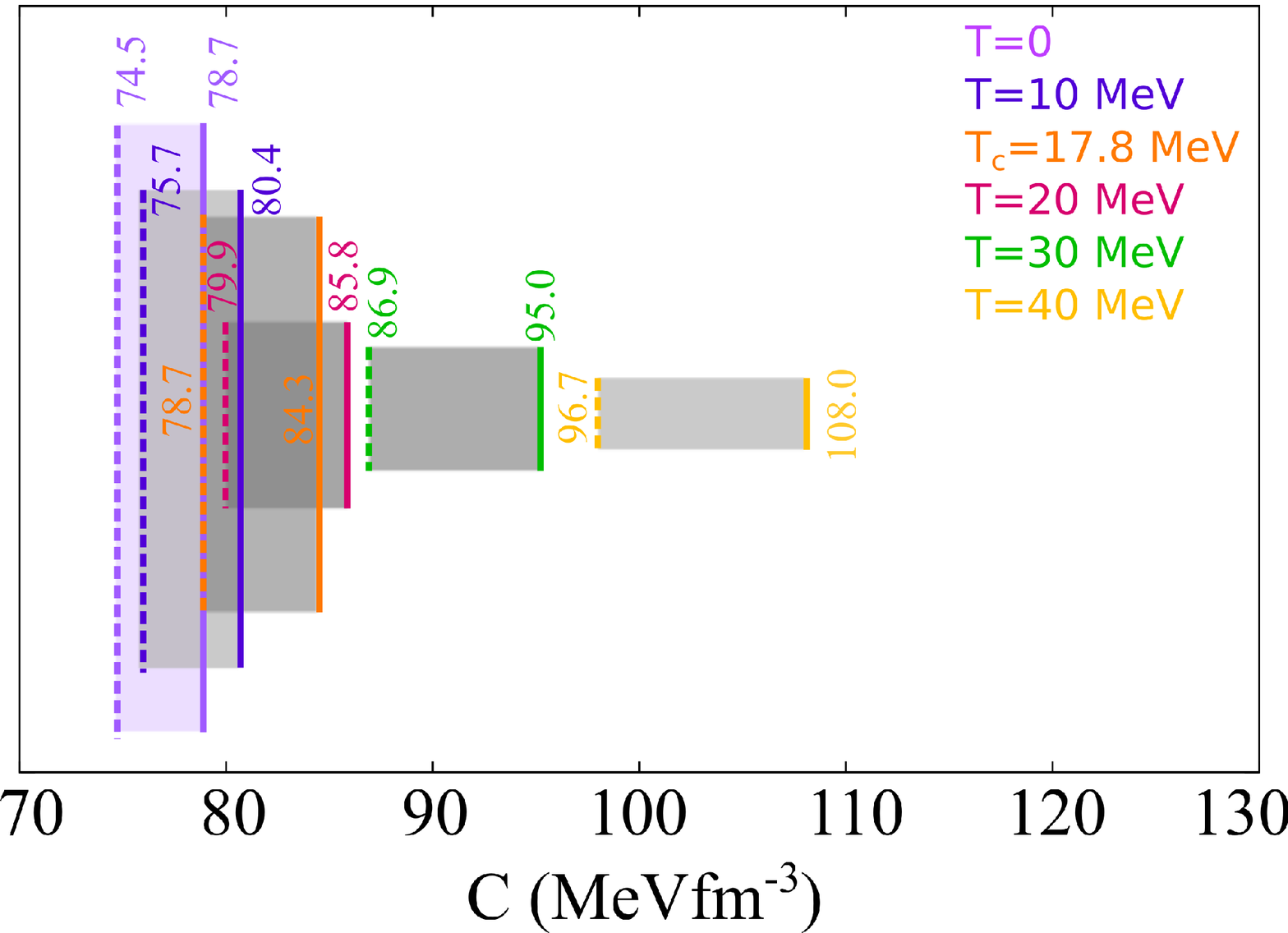}
 \includegraphics[width=0.45\textwidth]{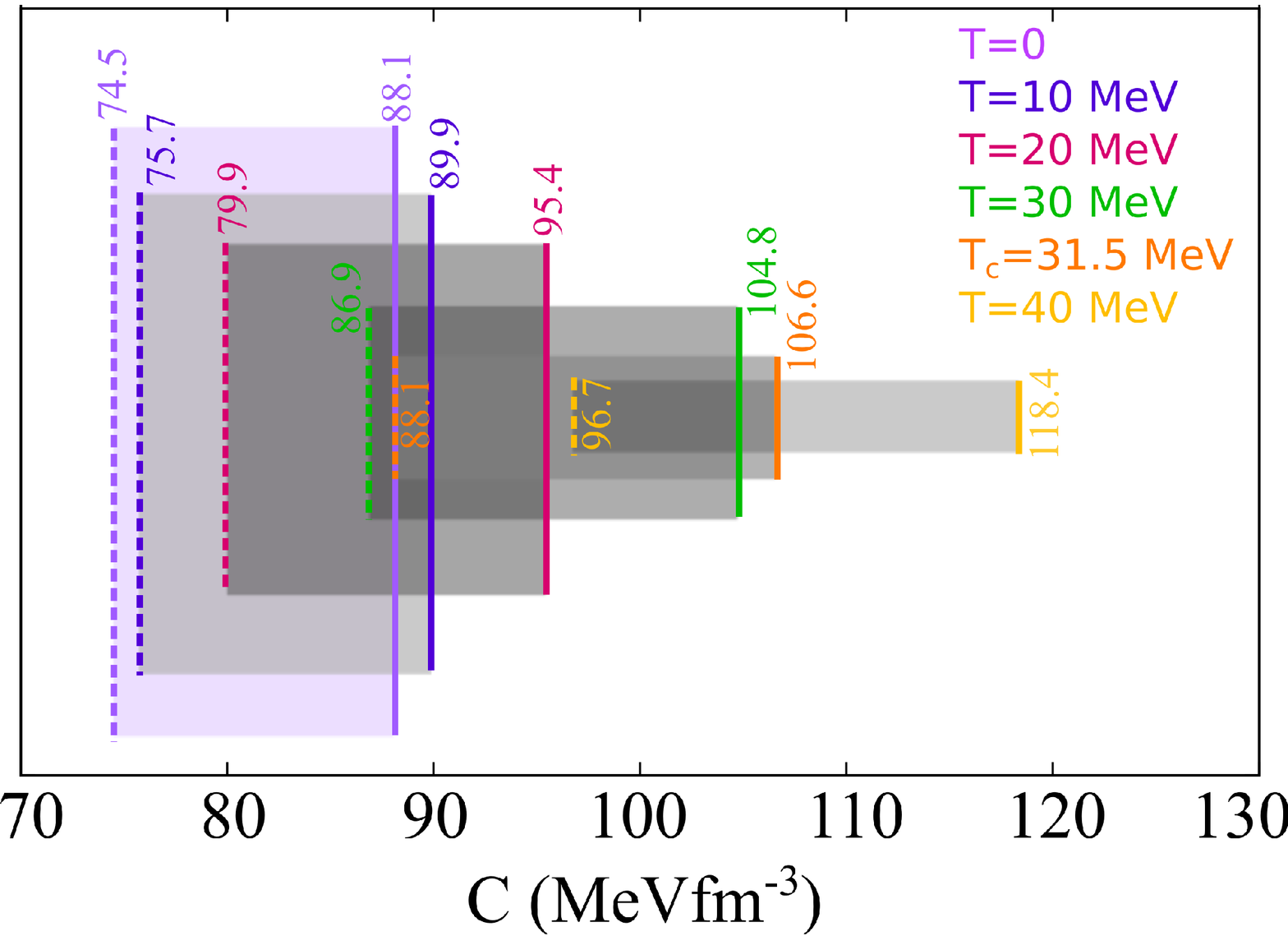}
\caption{(Color online) Stability windows of SM obtained with the QMDDv1 model with parametrizations B (top) and C (bottom) without the inclusion of magnetic field effects shown for different temperatures. The orange rectangle shows the highest temperature that overlaps with the zero temperature one.}
\label{box_stability_qmdd}
\end{figure}

In Fig.~\ref{box_stability_qmdd} the same calculation is performed for the QMDDv1 model using parametrizations B and C from Table~\ref{tab:table1}. 
In the first case, one can see that for $T=30$ MeV, no overlapping would be possible within this model if the lower limit of the stability window were
obtained also considering the finite temperature system.
This happens because there is an overall smaller overlap of stability windows for finite temperature with the QMDD than with the MIT bag model.
The last temperature where some overlap still exists with the zero temperature box is obtained for $T=17.1$ MeV for parametrization B and $T=31.7$ MeV for parametrization C. 

\begin{figure}[t!]
\centering
 \includegraphics[width=0.45\textwidth]{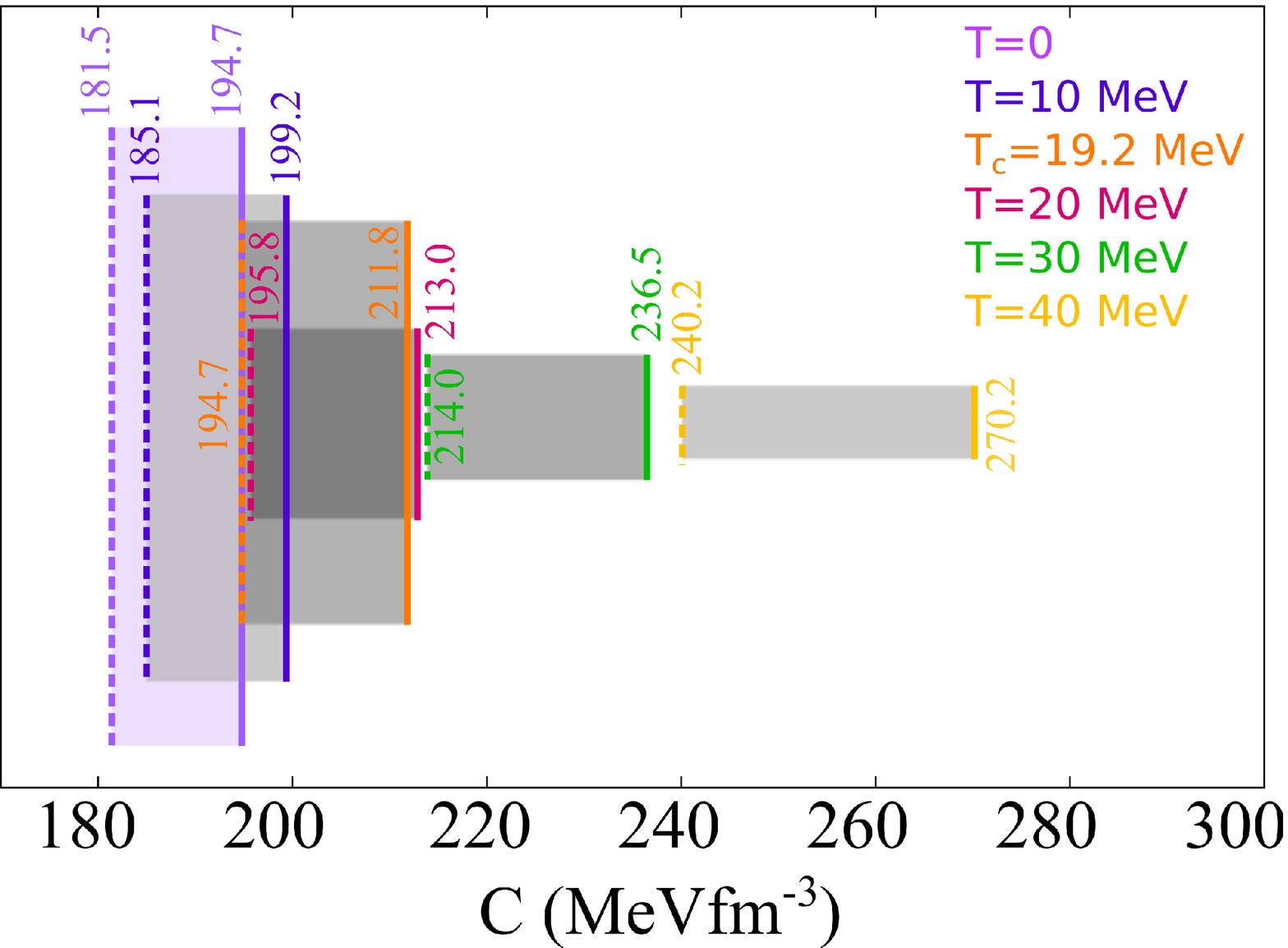}
 \includegraphics[width=0.45\textwidth]{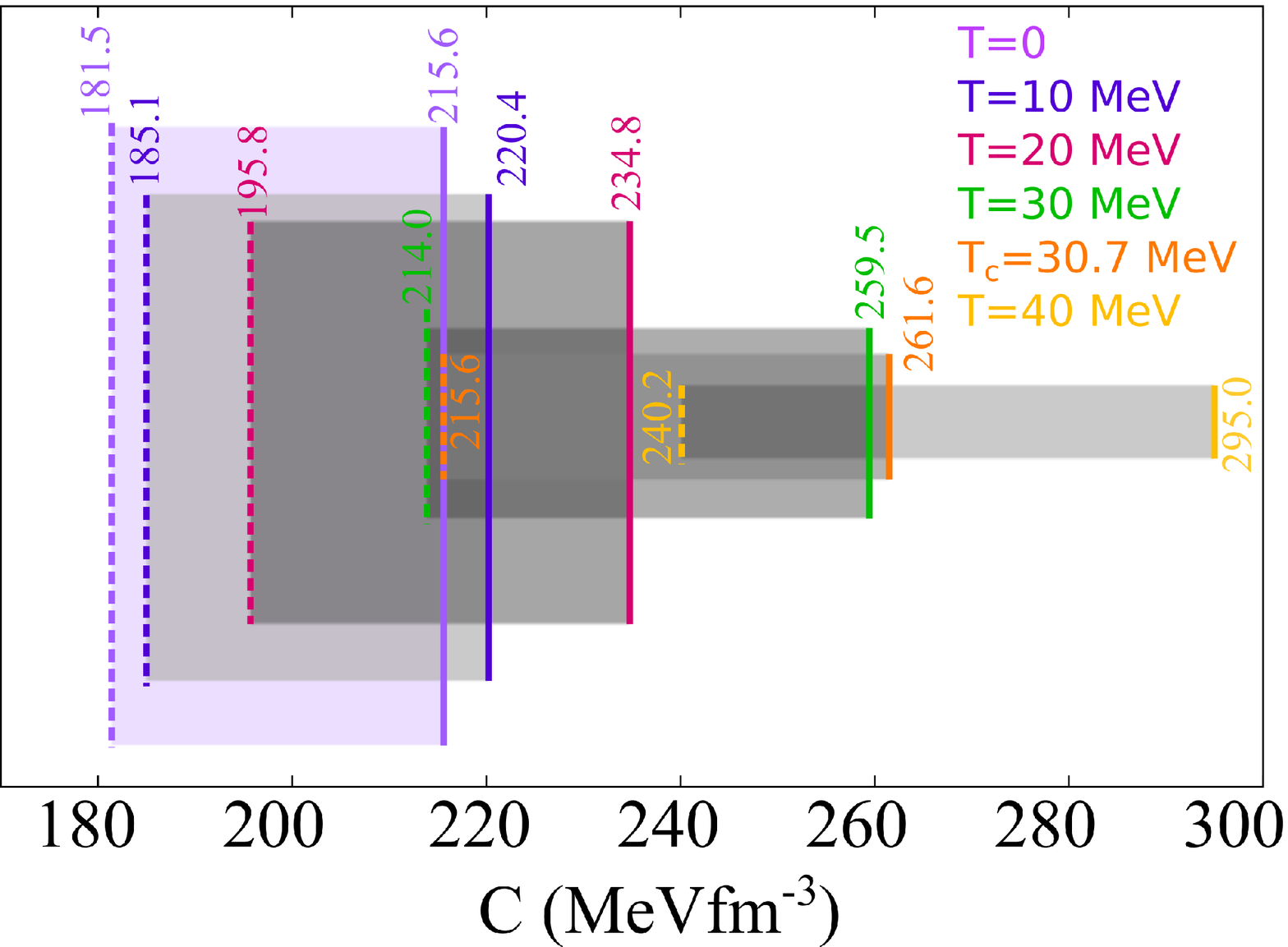}
\caption{(Color online) Stability windows of SM obtained with the QMDDv2 model with parametrizations D (top) and E (bottom) without the inclusion of magnetic field effects shown for different temperatures. The orange rectangle shows the highest temperature that overlaps with the zero temperature one.}
\label{box_stability_ddqm}
\end{figure}

For the sake of completeness, we recalculate the stability windows for the QMDDv2 model and the results are shown in 
 Fig.~\ref{box_stability_ddqm}. The interpretation of the appropriate parameter values for stellar matter follow the explanation given above.
Notice that, as expected from the results displayed in Fig.~\ref{box_stability_ddqm}, the overall parameter values are very different, making the equations of state of QMDDv2 much softer than the ones of QMDDv1. Comparing Figs. \ref{box_stability_qmdd}  and \ref{box_stability_ddqm}, one
can see that the maximum values allowed for the parameter $C$ within the QMDDv1 for $T=40$ MeV are much lower than
the minimum values allowed for the QMDDv2 model at zero temperature.

From hereon all calculations for the MIT bag model and the QMDD model are made considering the sets of parameters shown in Table~\ref{tab:table1} and discussed above. These parametrizations have realistic strange quark masses and they are the ones that comprehend stability windows for the largest possible range of temperatures.

\begin{figure}[t!]
\centering
 \includegraphics[width=0.5\textwidth]{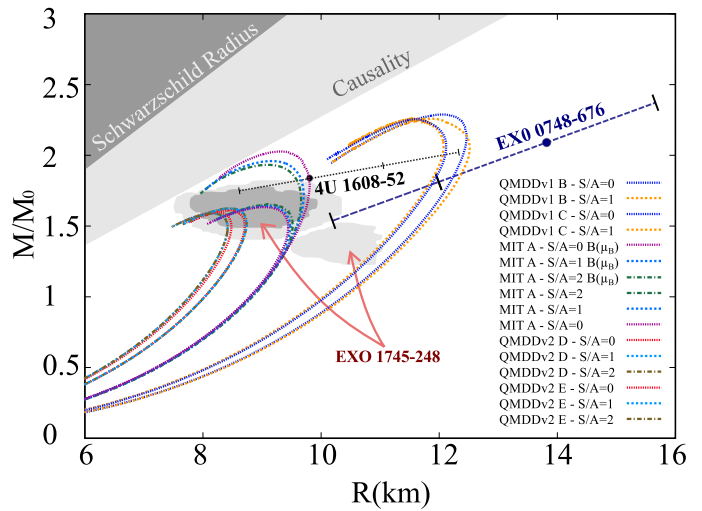}
\caption{(Color online) Mass-radius relation for different models shown for different entropies per baryon. We have also included lower and upper limits of the masses and radii of EXO 0748-676 and 4U 1608-52. The shaded clouds refer to the $1 \sigma$ and $2 \sigma$ confidence ellipse of the results obtained in for the EXO 1745-248 \cite{ozel1,ozel2,ozel3}. The four different sets of curves are presented in the legend going from higher star masses to lower star masses.}
\label{relacao_massa_raio}
\end{figure}

Finally, we have used the EOS's calculated with the parameters shown in Table~\ref{tab:table1} as input to the TOV equations \cite{tov1,tov2}. 
As already pointed out in many of the papers cited in out Introduction and specially in \cite{PerezMartinez:2005av, laura, Isayev:2011ug, Isayev:2013sq, nosso}, strong magnetic fields generate anisotropy in the pressure and in this case, the stiffness of the EOS is controlled by the
two different pressures and by the dependence of the magnetic field with the baryonic density. However, up to certain values of the magnetic
field, the splitting between the transverse and the parallel pressures is not substantial and hence, the use of the TOV equations  with
an isotropic EOS as input can be considered a valid approximation.
In this work we use the perpendicular pressure as the input for the TOV equations, as a first approximation, since our goal is only to compare different models. The resulting mass-radius relations are plotted in Fig. \ref{relacao_massa_raio}. The first model analyzed is the MIT at different entropies per baryon, indicating different evolution stages of the star without and with magnetic field effects.  In the cases in which the magnetic field is taken into account, it is considered to be variable, i.e., its magnitude increases with baryon chemical potential (density) following 
Refs.~\cite{Bandyopadhyay:1997kh,Mao:2001fq,Rabhi:2009ih,Dexheimer:2011pz}) from a surface value lower of $10^{15}$ G to a value lower than $7.2 \times 10^{18}$ G in the center of the star. Additionally, for this specific calculation, we add the pure magnetic field contribution proportional to $B^2$ to the pressure and energy density. As seen in many papers (as an example, see Refs.~\cite{luiz,veronica}), the magnetic field makes the EOS stiffer (when the pressure perpendicular to the magnetic field is used as input) and consequently, more massive stars are obtained in this case. This result is consistent with results from realistic simulations including magnetic fields \cite{Bocquet:1995je,Cardall:2000bs}. Note that when we include magnetic fields, the effect of a higher entropy per baryon (and consequently temperature) is to lower the maximum mass a star can have, instead of increasing it due to thermal effects. As pointed out in Ref.~\cite{veronica}, this is due to change in baryon number between 
different evolution stages, only allowed in stars belonging to binary systems.

\begin{table*}[t!]
\begin{ruledtabular}
\begin{tabular}{ccccccccc}
 Set & Model & $S/A$ & $Y_l$ & $B_c$ ($10^{18}$ G) & $M_{max}$ (M$_{\odot}$) & $R$ (km) & $\epsilon_c$ (fm$^{-4}$) & $T_c$ (MeV) \\ \hline
 A   & MIT   & $0$   & $N/F$ & $0$                 &$1.62$                   & $9.01$   & $8.25$                   & $0$         \\
 A   & MIT   & $1$   & $0.4$ & $0$                 &$1.64$                   & $9.10$   & $7.96$                   & $13.32$     \\
 A   & MIT   & $2$   & $0.4$ & $0$                 &$1.65$                   & $9.15$   & $7.85$                   & $26.59$     \\
 A   & MIT   & $0$   & $N/F$ & $6.64$              &$2.02$                   & $9.04$   & $8.31$                   & $0$         \\
 A   & MIT   & $1$   & $0.4$ & $4.71$              &$1.95$                   & $9.05$   & $8.82$                   & $12.20$     \\
 A   & MIT   & $2$   & $0.4$ & $4.44$              &$1.93$                   & $9.08$   & $8.69$                   & $24.27$     \\
\hline
 B   & QMDDv1  & $0$   & $N/F$ & $0$                 &$2.28$                   & $12.05$  & $4.56$                   & $0$         \\
 B   & QMDDv1  & $1$   & $0.4$ & $0$                 &$2.31$                   & $12.16$  & $4.39$                   & $11.34$     \\
 B   & QMDDv1  & $2$   & $0.4$ & $0$                 &$2.33$                   & $12.19$  & $4.46$                   & $22.80$     \\
 C   & QMDDv1  & $0$   & $N/F$ & $0$                 &$2.26$                   & $11.76$  & $4.65$                   & $0$         \\
 C   & QMDDv1  & $1$   & $0.4$ & $0$                 &$2.28$                   & $11.75$  & $4.76$                   & $11.78$     \\
 C   & QMDDv1  & $2$   & $0.4$ & $0$                 &$2.29$                   & $11.76$  & $4.79$                   & $23.62$     \\
 \hline
 D   & QMDDv2  & $0$   & $N/F$ & $0$                 &$1.60$                   & $8.42$   & $9.37$                   & $0$         \\
 D   & QMDDv2  & $1$   & $0.4$ & $0$                 &$1.62$                   & $8.46$   & $9.23$                   & $13.80$     \\
 D   & QMDDv2  & $2$   & $0.4$ & $0$                 &$1.62$                   & $8.46$   & $9.23$                   & $27.64$     \\
 E   & QMDDv2  & $0$   & $N/F$ & $0$                 &$1.59$                   & $8.22$   & $9.75$                   & $0$         \\
 E   & QMDDv2  & $1$   & $0.4$ & $0$                 &$1.58$                   & $8.16$   & $9.93$                   & $14.24$     \\
 E   & QMDDv2  & $2$   & $0.4$ & $0$                 &$1.58$                   & $8.16$   & $9.92$                   & $28.51$     \\
 
\end{tabular}
\end{ruledtabular}
\caption{\label{tab:table3} Output results given by the numerical solution of Tolman Oppenheimer Volkoff (TOV) equations for the MIT and QMDD models at different snapshots of the proto-quark star evolution. For the zero temperature case, the lepton fraction is not fixed (N/F).}
\end{table*}

For the QMDDv1 and QMDDv2 models, the results for different parametrizations are also shown in Fig. \ref{relacao_massa_raio}. 
While QMDDv1 always reproduces more massive stars than the MIT bag model due to the extra terms appearing in the grand-canonical thermodynamical potential, QMDDv2 models present results that are similar to the MIT ones, since the EOSs are much softer than the ones for QMDDv1.
One can see that the QMDDv1 model can reproduce very massive stars, as the ones recently measured \cite{demorest,antoniadis}, but fails to describe stars with low radii, if their existence is confirmed. On the other hand, the QMDDv2 model can describe stars with radii even lower than the MIT model.
 For different values of the confinement constant, it was already shown in Ref.~\cite{james} that this version of the QMDDv1 model generates larger maximum masses thant the MIT model systematically, for all possible parameters inside the stability window.
These results are not changed even when strong magnetic fields are considered in the MIT model.

Furthermore, we include fixed lepton fraction in the EOS's in order to improve our description of proto-quark stars. For the QMDD models, only electrons and their related neutrinos are considered within the lepton fraction, while in the MIT model, muons and their neutrinos are also taken into account.
The results for the QMDD and MIT bag model are shown in Table~\ref{tab:table3}. For the latter, we also show the inclusion of (variable) magnetic field effects. Note that the central magnetic field reached within the MIT bag model is higher at later stages of the star evolution, when the stars are also more massive due to a higher baryon number. In this table, besides maximum mass and respective radius, we also show central energy density and central temperature for each case considered.

\begin{figure}[t!]
\centering
 \includegraphics[width=0.45\textwidth]{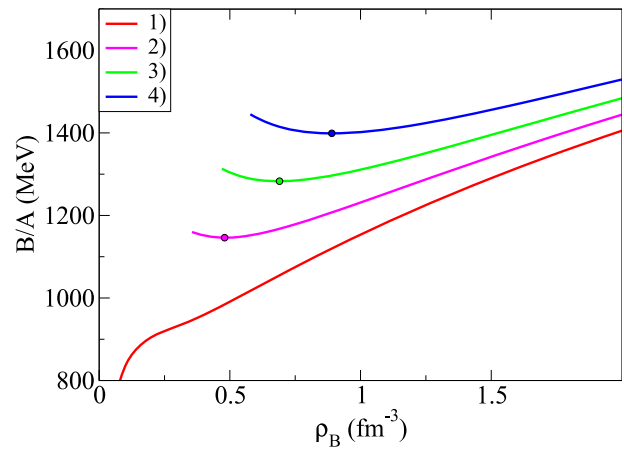}
\caption{(Color online) Binding energy per baryon as a function of baryon density shown for different parametrizations of the SU(2) NJL model at zero temperature. The dots are points of zero pressure.}
\label{njl1}
\end{figure}

We now analyze the  NJL model, for which the parameters are fixed so as to fit the values in vacuum for the pion mass, the pion decay constant,  the kaon mass and the quark condensates. We consider the sets of parameters given in Table
\ref{table2}, taken from Ref.~\cite{buballa} and references therein. In Fig. \ref{njl1}, the results for the SU(2) version of the model are shown. All four parametrizations are considered for $T=0$ and non-magnetized matter. We notice that parametrization 1 results in an unbound system. The other three parametrizations yield  bound systems, but the binding energies are larger than the one for iron. Therefore, the SU(2) version of the NJL model cannot be stable without the inclusion of the strange quark, as expected. Next, we include the strange quark in the calculation.

\begin{figure}[t!]
\centering
 \includegraphics[width=0.45\textwidth]{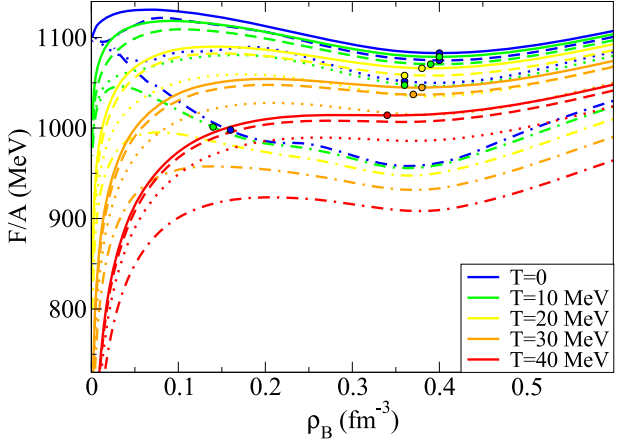}
\caption{(Color online) Free energy per baryon as a function of baryon density for parametrization set 5 of the SU(3) NJL model shown for different temperatures. Full lines represent calculations with $B=1\times 10^{18}$ G, dashed lines calculations with $B=5\times 10^{18}$ G, dotted lines $B=1\times 10^{19}$ G and dot-dashed lines $B=2\times 10^{19}$ G. The dots are points of zero pressure.}
\label{njl2}
\end{figure}

\begin{figure}[t!]
\centering
 \includegraphics[width=0.45\textwidth]{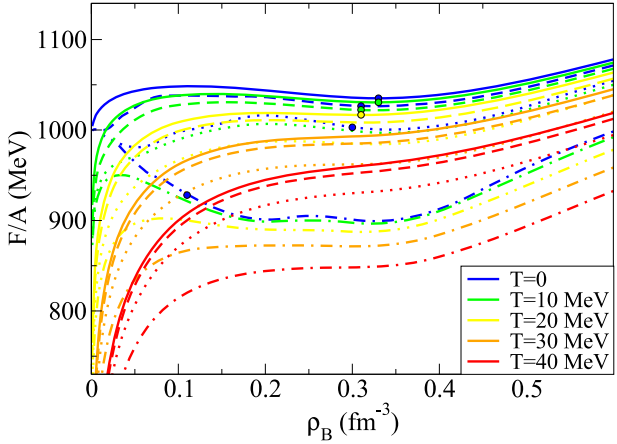}
\caption{(Color online) Free energy per baryon as a function of baryon density for parametrization set 6 of the SU(3) NJL model shown for different temperatures. Full lines represent calculations with $B=1\times 10^{18}$ G, dashed lines calculations with $B=5\times 10^{18}$ G, dotted lines $B=1\times 10^{19}$ G and dot-dashed lines $B=2\times 10^{19}$ G. The dots are points of zero pressure.}
\label{njl3}
\end{figure}

The results for both SU(3) parametrizations 5 and 6, shown in Table~\ref{table2} are seen in Figs.~\ref{njl2} and \ref{njl3}, respectively. 
A detailed discussion on the possibility of stable strange quark matter with the NJL model at zero temperature was done in section 3.3.2 of Ref.~\cite{buballa}, where the author claims that the NJL model does not support the idea of absolutely stable quark matter. However, in the present
work we can see that 
as the temperature increases, the free energy per baryon decreases, but the system becomes unbound. On the other hand, if we consider the influence of the magnetic field, the system becomes more bound as the magnetic field strength increases and, therefore, more stable. The latter had already been pointed out in 
Refs.~\cite{Anand:1999xx,Chakrabarty:1996te,Felipe:2008cm} for zero temperature. The dots show the points where the pressure is zero. These points do not coincide with the minima of the free energy per baryon for large magnetic fields.

\begin{table}[t!]
\begin{ruledtabular}
\begin{tabular}{ccccccc}
 Set & Group & $\Lambda$ [MeV] & G$\Lambda^2$ & $m_{u,d}$ [MeV] & K$\Lambda^5$  & $m_s$ [MeV] \\
\hline
 1   & SU(2) & 664.3           & 2.06         & 5               & N/A           & N/A         \\
 2   & SU(2) & 587.9           & 2.44         & 5.6             & N/A           & N/A         \\
 3   & SU(2) & 569.3           & 2.81         & 5.5             & N/A           & N/A         \\
 4   & SU(2) & 568.6           & 3.17         & 5.1             & N/A           & N/A         \\
\hline
 5   & SU(3) & 602.3           & 1.835        & 5.5             & 12.36         & 140.7       \\
 6   & SU(3) & 631.4           & 1.835        & 5.5             & 9.29          & 135.7       \\
\end{tabular}
\end{ruledtabular}
\caption{\label{table2}Parameters for the NJL model obtained from Ref. \cite{buballa} and references therein.}
\end{table} 

Note that for strong magnetic fields and low temperatures we see metastable configurations in Figs.~\ref{njl2} and \ref{njl3}. Those are related to the Haas-van Alphen oscillations \cite{a1,a2,a3,a4} that usually appear within these limits. These oscillations are related to when the Fermi energies of the charged particles cross the discrete threshold of a Landau level. See Ref.~\cite{Ferrari:2012yw} for more details on the effect of magnetic field on the Haas-van Alphen oscillations within the NJL model.

\section{Final remarks}

We presented in this paper a review of stability windows for quark matter at finite temperature using different models. We pointed out that the correct quantity to analyze in order to obtain the upper limits of the stability windows is the free energy per baryon, instead of the binding energy per baryon. In this case, for the QMDD as well as for the bag model, the correct values of the respective coupling constants (C and $\mathcal{B}^{1/4}$) assume larger values at larger temperatures. Even when a strong magnetic field is applied to the system, this behavior does not change, but it is enhanced.

Based on the results stated above, we chose various parametrizations for the QMDD model and two parametrizations for the bag model. For the latter they correspond to the cases without and with magnetic field effects. With these EOS's in hand, we simulate proto-quark stars for fixed  entropies per baryon and fixed lepton fraction. We see that in all cases analyzed (without and with magnetic field effects), the QMDDv1 reproduces more massive stars than the MIT bag model, while QMDDv2 produces stars with low masses and low radii.

The NJL model allows a more realistic description of quark matter, as it contains chiral symmetry restoration/breaking. For these reason, we studied both versions of the model, SU(2) and SU(3) in detail to check whether the stability for the existence of SM is satisfied with the parametrizations normally used in the literature.
The first group is not stable, as it is not bound below $930$ MeV. The second group has stable parametrizations for different temperatures (found by the analysis of the free energy per baryon), specially when strong magnetic fields are considered.

\section*{Acknowledgements} 
This work was partially supported by CNPq and CAPES. The authors acknowledge fruitful discussions with Prof. S\'ergio Barbosa Duarte and Marcus Benghi Pinto.

\section*{Appendix}

Here we present the most relevant formulas for the NJL model without the inclusion of magnetic field effects. Once more, we start from Eq. (\ref{thermopot}), from which we obtain for the energy density and entropy density of the system in the mean field approximation 
\begin{eqnarray}
{\epsilon}\,& =&-2 N_c \sum_i \int \frac{d^3k_i}{(2\pi)^3} 
\frac{k_i^2 +  m_i M_i}{E_i} \,(f_{- i} - f_{+ i}) \theta(\Lambda^2-k_i^2)
\nonumber\\
&-&2G(\phi_u^2+\phi_d^2+\phi_s^2) + 2K \phi_u \phi_d \phi_s -{\epsilon}_0 \mbox{,}\label{enjl}
\end{eqnarray}
\begin{eqnarray}
s&=&-2 N_c \sum_i \int \frac{d^3k_i}{(2\pi)^3}  \Big[f_{+i} \ln\left(f_{+i}\right)
+\left(1-f_{+ i}\right)  \nonumber\\
&\times&\ln\left(1-f_{+ i}\right)+f_{- i} \ln\left(f_{- i}\right)
+\left(1-f_{- i}\right) \ln\left(1-f_{- i}\right)\Big] \nonumber \\
& \times&\theta(\Lambda^2-k_i^2) \mbox{,} 
\end{eqnarray}
where $E_i^*=\sqrt{k_i^2+M_i^2}$ and $M_i$ is the constituent mass of each quark.  Minimizing the grand-canonical thermodynamical potential $\Omega$ with respect to $M_i$ leads to the three gap equations Eq.~(\ref{efmass}). The baryon density has the same form as given in Eq.~(\ref{densidade_quarks_qmdd}).

The quark condensates are defined, for each of the flavors $i=u,d,s$  as
\begin{equation}
\phi_i=\langle\bar q_i\, q_i\rangle = 
-2 N_c \sum_i \int \frac{d^3k_i}{(2\pi)^3} 
\frac{M_i}{E_i} 
(\,f_{+ i}\,-\,f_{- i}\,) \theta(\Lambda^2-k_i^2)
 \mbox{.} \label{qcond}
\end{equation}


\section*{References}

\begin{thebibliography}{99}

\bibitem{Lattimer2004} J. M. Lattimer and M. Prakash, Science {\bf 304}, 
536 (2004).

\bibitem{prak97} M. Prakash, I. Bombaci, M. Prakash, P. J. Ellis, J. M.
Lattimer and R. Knorren, Phys. Rep. {\bf 280}, 1 (1997).

\bibitem{Glen00} N. K. Glendenning, Compact Stars, Springer-Verlag, New-York,
2000.

\bibitem{pons2001} J.A. Pons, A.W. Steiner, M. Prakash and J.M. Lattimer,
Phys.Rev.Lett. {\bf 86} (2001) 5223-5226.

\bibitem{pons1999} J.A. Pons, S. Reddy, M. Prakash, J.M. Lattimer and J.A. 
Miralles, Astrophys. J. {\bf 513}, 780 (1999).

\bibitem{marcelo} M. D. Alloy and D. P. Menezes, Phys. Rev. C {\bf 83}, 035803 (2011).

\bibitem{Itoh} N. Itoh, Prog. Theor. Phys., 44, 291-292 (1970).

\bibitem{bodmer} A.R. Bodmer, Phys. Rev. D {\bf 4}, 1601 (1971).

\bibitem{witten}  E. Witten, Phys. Rev. D \textbf{30}, 272 (1984).

\bibitem{olinto} C. Alcock, E. Farhi and A. Olinto, Astrophys. J. {\bf 310},
261 (1986).

\bibitem{mit1} A. Chodos, R.L. Jaffe, K. Johnson, C.B. Thorne and V.F. 
Weisskopf, Phys. Rev. D \textbf{9}, 3471 (1974).

\bibitem{mit2}E. Farhi and R.L. Jaffe, Phys. Rev. D \textbf{30}, 11 (1984).

\bibitem{njl} Y. Nambu and G. Jona-Lasinio, Phys. Rev. {\bf  122}, 345 (1961).

\bibitem{njl2}  Y. Nambu and G. Jona-Lasinio, Phys. Rev. {\bf 124}, 246 (1961).

\bibitem{hybrid2}M.G. Paoli and D.P. Menezes, Eur. Phys. J. A {\bf 46}, 413 (2010).

\bibitem{hybrid4}D.P. Menezes and C. Provid\^encia, Phys. Rev. C
{\bf 69}, 045801 (2004).

\bibitem{hybrid6}D.P. Menezes and C. Provid\^encia, 
Phys. Rev. C {\bf 68}, 035804 (2003).

\bibitem{Alford:2007xm} 
  M.~G.~Alford, A.~Schmitt, K.~Rajagopal and T.~Schafer,
  Rev.\ Mod.\ Phys.\  {\bf 80}, 1455 (2008).

\bibitem{sergio1} J. C. Oliveira, H. Rodrigues, and S. B. Duarte,
  Phys. Rev. D {\bf 78}, 123008 (2008).

\bibitem{sergio2} M. Orsaria, H. Rodrigues and
  S.B. Duarte, Int. J. Mod. Phys. D {\bf 16}, 291 (2007).

\bibitem{lugones2010} G. Lugones, T.A.S. do Carmo, A.G. Grunfeld
and N.N. Scoccola, Phys. Rev. D {\bf 81}, 085012 (2010).

\bibitem{laura} L. Paulucci, E.J. Ferrer, V. de la Incera and J.E. Horvath,
Phys. Rev. D {\bf 83}, 043009 (2011).

\bibitem{fowler} G.N. Fowler, S. Raha and R.M. Weiner, Z. Phys. {\bf C 9}, 271 (1981).

\bibitem{sedrakian2011} L.Bonanno and A. Sedrakian, Astron. \& Astrophys. 539, A16 (2012).

\bibitem{novos} D. Nickel, Phys. Rev. Lett. {\bf 103}, 072301 (2009)

\bibitem{novos2} M. Buballa and D. Nickel, Acta Phys. Polon. Supp. {\bf 3},
523 (2010)

\bibitem{novos3} T.~Kojo, Y.~Hidaka, L.~McLerran and R.~D.~Pisarski,
  Nucl.\ Phys.\ A {\bf 843}, 37 (2010).

\bibitem{lugones95} O.G. Benvenuto and G. Lugones G, Phys. Rev. D \textbf{51}, 1989 (1995).

\bibitem{1995NuPhA.588..365T} T.~Takatsuka, Nucl. Phys. A {\bf 588}, 365 (1995).

\bibitem{Prakash:1996xs} M.~Prakash, I.~Bombaci, M.~Prakash, P.~J.~Ellis, J.~M.~Lattimer and R.~Knorren,
 Phys. Rept.  {\bf 280}, 1 (1997).

\bibitem{james} 
  J.~R.~Torres and D.~P.~Menezes,
  Europhys.\ Lett.\  {\bf 101}, 42003 (2013).

\bibitem{nucleation1} O. G. Benvenuto and G. Lugones, Mon. Not. R. Astron.
Soc. {\bf 304}, L25 (1999).

\bibitem{nucleation2}I. Bombaci, D. Logoteta, P. K. Panda, 
C. Providencia, and I. Vidana, Phys. Lett. B {\bf 680}, 448 (2009).

\bibitem{Dexheimer:2008ax} V.~Dexheimer and S.~Schramm,
 Astrophys. J.  {\bf 683}, 943 (2008).

\bibitem{veronica} V. Dexheimer, D.P. Menezes and M. Strickland,
arXiv:1210.4526 [nucl-th].

\bibitem{Li:2013xv} 
  Z.~Li,
  arXiv:1302.0104 [astro-ph.SR].

\bibitem{Li:2007kg} 
  Z.~Li and Z.~Han,
  Astrophys. J.  {\bf 685}, 225 (2008).

\bibitem{chmaj89} T. Chmaj and W. Slominski, Phys. Rev. D {\bf 40},
  165 (1988).

\bibitem{chakrabarty93} S. Chakrabarty, Phys. Rev. D {\bf 48}, 1409 (1993).

\bibitem{lugones95t} G. Lugones and O.G. Benvenuto, Phys. Rev. D {\bf 52}, 1276 (1995).

\bibitem{su2002} Y. Zhang and R-K Su, Phys. Rev. C {\bf 65}, 035202
  (2002).

\bibitem{greiner} W. Greiner, L. Neise and H. Stocker, {\it
    Thermodynamics and Statistical Mechanics}, Springer, 1997.

\bibitem{pastat1} S.S. Avancini, C.C.  Barros, D.P. Menezes and
  C. Providência, Phys. Rev. C {\bf 82}, 025808 (2010).

\bibitem{pastat2}S.S. Avancini, D.P. Menezes, M.D. Alloy, J.R. Marinelli, M.M.W. de Moraes 
and C. Provid\^encia,  Phys. Rev. C {\bf 78}, 015802 (2008). 

\bibitem{c11} B.~Paczynski, Acta Astron.\ {\bf 42}, 145 (1992).

\bibitem{c12}C.~ Thompson and R.~C.~Duncan, Astrophys.\ J. {\bf 392}, L9 (1992).

\bibitem{c13}C.~ Thompson and R.~C.~Duncan, Astrophys.\ J. {\bf 473}, 322 (1996).

\bibitem{c14}A.~Melatos, Astrophys.\ J.\ Lett. {\bf 519}, L77 (1999).

\bibitem{b1}
J.~H.~Taylor, R.~N.~Manchester, and A.~G.~Lyne,
Astrophys.\ J. S {\bf 88}, 529 (1993).

\bibitem{Ferrer:2010wz} 
  E.~J.~Ferrer, V.~de la Incera, J.~P.~Keith, I.~Portillo and P.~P.~Springsteen,
  Phys.\ Rev.\ C {\bf 82}, 065802 (2010).


\bibitem{PerezMartinez:2005av} 
  A.~Perez Martinez, H.~Perez Rojas, H.~J.~Mosquera Cuesta, M.~Boligan and M.~G.~Orsaria,
  Int.\ J.\ Mod.\ Phys.\ D {\bf 14}, 1959 (2005).

\bibitem{Felipe:2007vb} 
  R.~G.~Felipe, A.~P.~Martinez, H.~P.~Rojas and M.~Orsaria,
  Phys.\ Rev.\ C {\bf 77}, 015807 (2008).

\bibitem{Chaichian:1999gd} 
  M.~Chaichian, S.~S.~Masood, C.~Montonen, A.~Perez Martinez and H.~Perez Rojas,
  Phys.\ Rev.\ Lett.\  {\bf 84}, 5261 (2000).

\bibitem{Martinez:2003dz} 
  A.~P.~Martinez, H.~P.~Rojas and H.~J.~Mosquera Cuesta,
  Eur.\ Phys.\ J.\ C {\bf 29}, 111 (2003).

\bibitem{PerezMartinez:2007kw} 
  A.~Perez Martinez, H.~Perez Rojas and H.~Mosquera Cuesta,
  Int.\ J.\ Mod.\ Phys.\ D {\bf 17}, 2107 (2008).

\bibitem{Orsaria:2010xx} 
  M.~Orsaria, I.~F.~Ranea-Sandoval and H.~Vucetich,
  Astrophys.\ J.\  {\bf 734}, 41 (2011).

\bibitem{Isayev:2011ug} 
  A.~A.~Isayev and J.~Yang,
  Phys.\ Lett.\ B {\bf 707}, 163 (2012).

\bibitem{Isayev:2013sq} 
    A.~A.~Isayev and J.~Yang,
  Phys.\ Rev.\ C {\bf 84}, 065802 (2011).

\bibitem{Strickland:2012vu} 
  M.~Strickland, V.~Dexheimer and D.~P.~Menezes,
  Phys.\ Rev.\ D {\bf 86}, 125032 (2012).
  
\bibitem{Sinha:2012cx} 
  M.~Sinha, B.~Mukhopadhyay and A.~Sedrakian,
  Nucl.\ Phys.\ A {\bf 898}, 43 (2013).

\bibitem{chakrabarty91} S. Chakrabarty, Phys. Rev. D \textbf{43}, 627 (1991).

\bibitem{wang00} P. Wang, Phys. Rev. C  {\bf 62}, 015204 (2000).

\bibitem{peng00} G.X. Peng, H.C. Chiang, B.S. Zou, P.Z. Ning and
  S.J. Luo, Phys. Rev. C {\bf 62}, 025801 (2000).

\bibitem{su2008} S. Yin and R-K. Su, Phys. Rev. C {\bf 77}, 055204
  (2008).

\bibitem{njlb1} D.P. Menezes, M. Benghi Pinto, S.S. Avancini, 
A. P\'erez Martinez and C. Provid\^encia, Phys. Rev. {\bf C 79}, 035807 (2009).

\bibitem{njlb2}S.S. Avancini, D.P. Menezes, M.B. Pinto and C. Provid\^{e}ncia,
Phys. Rev. {\bf C 80}, 065805 (2009).

\bibitem{njlb3}S.S. Avancini, D.P. Menezes, M.B. Pinto and C. Provid\^encia,
Phys. Rev. {\bf D 85}, 091901(R) (2012).

\bibitem{wen2005} X.J. Wen, X.H. Zhong, G.X. Peng, P.N. Shen and P.Z. Ning,
Phys. Rev {\bf C 72}, 015204 (2005).

\bibitem{Ebert:1999ht} 
  D.~Ebert, K.~G.~Klimenko, M.~A.~Vdovichenko and A.~S.~Vshivtsev,
  Phys.\ Rev.\ D {\bf 61}, 025005 (2000).

\bibitem{quarkionicas1} D.P. Menezes, C. Provid\^encia and D.B. Melrose,
 J. Phys. G: Nucl. Part. Phys. \textbf{32}, 1981 (2006).

\bibitem{Burrows:1986me} 
  A.~Burrows and J.~M.~Lattimer,
  Astrophys.\ J.\  {\bf 307}, 178 (1986).

\bibitem{tov1} Tolman, R.C., Phys. Rev. \textbf{55}, 364 (1939).

\bibitem{tov2} J.R. Oppenheimer and G.M. Volkoff,  Phys. Rev. \textbf{55}, 374
  (1939).

\bibitem{nosso} V. Dexheimer, D.P. Menezes and M. Strickland,
arXiv:1210.4526[nucl-th].

\bibitem{Bandyopadhyay:1997kh} 
  D.~Bandyopadhyay, S.~Chakrabarty and S.~Pal,
  Phys.\ Rev.\ Lett.\  {\bf 79}, 2176 (1997).

\bibitem{Mao:2001fq} 
  G.~-J.~Mao, A.~Iwamoto and Z.~-X.~Li,
  Chin.\ J.\ Astron.\ Astrophys.\  {\bf 3}, 359 (2003).

\bibitem{Rabhi:2009ih} 
  A.~Rabhi, H.~Pais, P.~K.~Panda and C.~Providencia,
  J.\ Phys.\ G G {\bf 36}, 115204 (2009).

\bibitem{Dexheimer:2011pz} 
  V.~Dexheimer, R.~Negreiros and S.~Schramm,
  Eur.\ Phys.\ J.\ A {\bf 48}, 189 (2012).

\bibitem{luiz}  L.~L.~Lopes and D.~P.~Menezes,
  Brazilian Journal of Physics 42 (2012).

\bibitem{Bocquet:1995je} 
  M.~Bocquet, S.~Bonazzola, E.~Gourgoulhon and J.~Novak,
  Astron.\ Astrophys.\  {\bf 301}, 757 (1995).

\bibitem{Cardall:2000bs} 
  C.~Y.~Cardall, M.~Prakash and J.~M.~Lattimer,
  Astrophys.\ J.\  {\bf 554}, 322 (2001).

\bibitem{demorest} P. B. Demorest, T. Pennucci, S. M. Ransom,
  M. S. E. Roberts \& J. W. T. Hessels, Nature 
\textbf{467}, (1081–1083) (2010).

\bibitem{antoniadis} J.~Antoniadis, P.~C.~C.~Freire, N.~Wex, T.~M.~Tauris, R.~S.~Lynch, M.~H.~van Kerkwijk, M.~Kramer and C.~Bassa {\it et al.},
  Science {\bf 340}, 6131 (2013).

\bibitem{ozel1} F. Ozel, G. Baym, and T. Guver, Phys. Rev. D {\bf 82}, 101301 (2010).

\bibitem{ozel3} F. \"Ozel, T. G\"uver, D. Psaltis, ApJ \textbf{693}, 1775 (2009).

\bibitem{ozel2}F. \"Ozel, Nature \textbf{441}, 1115 (2006).

\bibitem{buballa} M. Buballa, Phys. Rep. {\bf 407} (2005) 205.

\bibitem{Anand:1999xx} 
  J.~D.~Anand and S.~Singh,
  Pramana {\bf 52}, 127 (1999).

\bibitem{Chakrabarty:1996te} 
  S.~Chakrabarty,
  Phys.\ Rev.\ D {\bf 54}, 1306 (1996).

\bibitem{Felipe:2008cm} 
  R.~G.~Felipe and A.~P.~Martinez,
  J.\ Phys.\ G {\bf 36}, 075202 (2009).

\bibitem{a1}
W.~J.~de Haas and P.~M. van Alphen,
Leiden\ Commun.\ A {\bf 212}, 215 (1930).

\bibitem{a2}
W.~J.~de Haas and P.~M. van Alphen,
Proc.~R.~Acad.~Sci.~Amsterdam {\bf33}, 1106 (1930).

\bibitem{a3}
D.~Ebert, K.~G.~Klimenko, M.~A.~Vdovichenko and A.~S.~Vshivtsev,
Phys.\ Rev.\ D {\bf 61}, 025005 (2000).

\bibitem{a4}
D.~Ebert and K.~G.~Klimenko,
Nucl.\ Phys.\ A {\bf 728}, 203 (2003).


\bibitem{Ferrari:2012yw} 
  G.~N.~Ferrari, A.~F.~Garcia and M.~B.~Pinto,
  Phys.\ Rev.\ D {\bf 86}, 096005 (2012).

\end{thebibliography}
 \end{document}